# Finite temperature external potential in crystalline solids

T. R. S. Prasanna and M. P. Gururajan

Department of Metallurgical Engineering and Materials Science

Indian Institute of Technology, Bombay

Powai, Mumbai – 400076

India

Thermal vibrations alter the external potential. Allen (Phys. Rev. B 18 (1978) 5217) proved that at finite temperatures the pseudopotential form factors are corrected by a Debye-Waller Factor (DWF). We generalize this result to the crystal potential. (The generalization to the all-electron case of the nuclear potential fails due to the breakdown of the rigid-atom approximation.) This finite temperature formalism only gives thermal-averaged properties and no dynamical information can be obtained. Hence, it is labeled the Quasi Ab Initio formalism. Analogous to the use of experimental lattice parameters in ab initio studies, experimental DWF can also be used. The justification is identical; the experimental parameters can be validated by separate ab initio studies. Our work transforms, forty years later, Kasowski's empirical study (Phys. Rev. B. 8 (1973) 1378) into the first ab initio finite temperature band structure calculation. This formalism opens the way to obtain ab initio finite temperature thermal-averaged properties from a single calculation.

# 1. Introduction

In the basic Hamiltonian for crystalline solids, electrons move in an external potential due to fixed nuclear point charges in a static lattice at 0 K. The dominant method to incorporate the role of thermal vibrations at finite temperatures is through molecular dynamics. An alternate approach is the Allen-Heine theory [1-3] that has been used mostly to account for the temperature dependence of band gaps in semiconductors [4-9] and infrequently to explain other experimental observations. The self-energy contribution to the electron energy, due to thermal vibrations, was first suggested by Fan [10]. In contrast, starting with Antoncik [11] in the 1950s, several authors [4-7] have empirically applied a Debye-Waller Factor (DWF) correction to the potential to account for the temperature dependence of band gaps. Baumann [12] was the first to suggest that both self-energy and DWF corrections are necessary for a complete account of the role of thermal vibrations on electron energies. This approach was put on a rigorous footing by Allen and Heine [1] and was further developed by Allen and co-workers [2, 3].

The Allen-Heine (AH) Theory [1] is a 2$^{nd}$ order perturbation theory. However AH also hinted at a more accurate approach as "*A higher-order adiabatic perturbation summation can be accomplished by solving $H_0 + \overline{H}_2 + \overline{H}_4 + ...$ exactly and then using the resulting temperature-dependent eigenfunctions and energies to calculate the self-energy terms.*"

This more accurate approach was later developed for pseudopotentials by Allen [2]. It is especially useful at high temperatures since the Debye-Waller term is valid to all orders. However, it has not yet been implemented in any first principles calculations despite the absence of any theoretical or computational impediments. We discuss the reason for this neglect and show that it is unwarranted. In contrast, the less accurate Allen-Heine Theory [1] continues to be of interest [9, 13-17] as it is compatible with ab initio studies.

It is evident from Allen's proof [2] that valence electrons effectively move in the background of temperature dependent static lattice pseudopotential. The question of whether this result can be generalized beyond pseudopotentials has not been addressed till date. We address this question in the present article. We also discuss the possibilities and limitations of this formalism.

**2. Theory**

We begin by drawing attention to sec-4 and 5 of Allen [2]. The electron-phonon interaction Hamiltonian is given by Eq.17 of Allen [2] as

$$H_{e-p} = \sum_l [V(\vec{r} - \vec{R_l}) - V(\vec{r} - \vec{l})] \tag{A17}$$

where all terms have their usual meaning. While sec-4 of Allen [2] develops the second order perturbation theory, our interest is in the results obtained in sec-5 using a plane-wave basis set. Their advantage is described by Allen [2] as "*In a plane-wave basis set, all non-vanishing terms can be kept and the result is Debye-Waller corrections to the crystal potential.*" Following the derivation from Eq.27-Eq.30, the final result is given by Eq.30 of Allen [2] as

$$\langle H_{e-p} \rangle = \sum_{\vec{k}}^{BZ} \sum_{\vec{G}\vec{G'}} V(\vec{G'} - \vec{G})(e^{-W(\vec{G'}-\vec{G})} - 1) c^+_{\vec{k}+\vec{G'}} c_{\vec{k}+\vec{G}} \tag{A30}$$

Allen [2] states "*The perturbation, Eq.30, is now periodic and can be added onto the unperturbed Hamiltonian. The result is a reduction of the pseudopotential form factors V(G) by*

*Debye-Waller factors $e^{-W(G)}$.*" Importantly, the derivation relies on V(q), the Fourier Transform of the pseudopotential, and not just on V(G) that is defined only on reciprocal lattice vectors for empirical pseudopotentials. Thus, Eq.A30 can also be used with ab initio pseudopotentials.

The assumptions underlying Allen's derivation [2] of Eq.A30 are 1) adiabatic approximation and 2) the rigid atom approximation. It is evident from Allen's derivation [2] that the final result, Eq.A30, is independent of any specific feature of pseudopotentials. It immediately follows that Eq.A30 should be valid for any potential that satisfies the underlying assumptions.

**2.1 Extension to crystal potential**

The first case that we consider is the crystal potential that is the sum of the nuclear potential and the Coulomb potential due to electrons. The nuclear potential always moves rigidly with the nucleus since the nucleus is a point charge and hence satisfies the above assumptions. Therefore, if the rigid atom approximation is made for the potential due to electrons, the crystal potential moves rigidly with the atoms. Hence, Allen's result can be extended to the crystal potential. It follows that at finite temperatures the crystal potential form factor is modified by the Debye Waller factor and is given by

$$V_i^{cr}(\boldsymbol{G},T) = V_i^n(\boldsymbol{G},T) + V_i^e(\boldsymbol{G},T) = [V_i^n(\boldsymbol{G},0) + V_i^e(\boldsymbol{G},0)]e^{-W_i(\boldsymbol{G},T)} \qquad (3)$$

This opens up this formalism to electronic structure calculation methods that are not based on pseudopotentials. An example is discussed later.

## 2.2 Extension to nuclear potential

The second case that we consider is if this formalism can be extended to all electron methods where the external potential is just the nuclear potential.

As is well known, the number of plane waves needed for bare nuclear potentials make computations impractical. Pseudopotentials are necessary to reduce the number of plane waves. However, Eq.A30 is a theoretical result and computational considerations are secondary. There is no restriction on the number of terms (or plane waves) in Eq.A30 which, in principle, can be infinite. Thus, the potential in Eq.A17 can also be viewed to be solely due to the nuclear charges and the final result of Allen [2] using plane-wave basis set, Eq.A30, remains valid. Therefore, the crystal potential in Eq.3 is given by just the nuclear potential term in the all electron case.

It is well known that the DWF transforms the nuclear point charge into a Gaussian distribution [18]. From the Poisson Equation, the thermal-averaged nuclear potential is obtained as [18]

$$V_i^n(r,T) = \frac{Z_i e}{r} \, erf\left[\frac{r}{(2\langle u_i^2(T)\rangle)^{1/2}}\right] \tag{4}$$

The above potential implies that all electrons move in the background of a thermal-averaged nuclear potential at finite temperatures. Therefore, the solution of the atomic Schrodinger equation using Eq.4 will give the finite temperature core states.

We have performed this exercise for Si using the code RAtom [19]. Specifically, in this code we have replaced the Ze/r nuclear potential by the modified potential, Eq.4. Using $\langle u_i^2(T)\rangle$ to be zero for the T = 0 K case, the modified code gives energies that match the energies reported in the NIST Database [20]. For other temperatures, ab initio $\langle u_i^2(T)\rangle$ values for Si reported in Table 6 of Ref.21 were used.

Table 1 gives the energy levels for the occupied states for both the singular and the thermal-averaged (Eq.4) nuclear potential. It is seen from Table 1 that the core state energies vary drastically at finite temperatures. However, it is well known that the core state energies are independent of temperature. This indicates that the extension of Allen's formalism to the case where the external potential is the nuclear potential is incorrect.

The reason for the failure lies in the breakdown of the rigid atom approximation. The basis for the rigid atom approximation is that the nuclear motion is much slower than electron motion. When the nucleus moves, the electrons are in equilibrium at any position of the nucleus and the electronic energy levels remain unchanged, especially for core electrons. But it is seen from Table 1 that core energy levels are temperature dependent and very different from the core energy levels in an atom.

Therefore, when the external potential is just the nuclear potential, the net result is that the rigid atom approximation is no longer valid for core electrons. The original derivation [2] of Eq.A30 was made under this assumption. Since the generalization to nuclear potential violates this assumption it is invalid. (The rigid atom approximation is central to the Allen-Heine Theory in other contexts as well [15].)

## 3. Discussion

The present work generalizes Allen's result [2] and extends it beyond pseudopotentials to the crystal potential. Taken together, there now exists a theoretical formalism to perform finite temperature electronic structure calculations. From the resulting wavefunctions and energies, self energy corrections can be calculated as discussed by Allen [2].

We first discuss the neglect of Allen's [2] result that is valid for pseudopotentials, including ab initio pseudopotentials. This neglect is not due to theoretical or computational reasons. We recall that the less accurate Allen-Heine theory [1], which does not use DWF, continues to be of interest since it is compatible with ab initio studies [9, 13-17]. Therefore, the most likely reason for this neglect of the more accurate theory of Allen [2] is conceptual *viz.* the impression that using experimental DWF makes the method empirical and not ab initio. We show below that this impression is unwarranted.

### 3.1 Experimental parameters are already in use in ab initio studies

Ab initio calculations are routinely performed using experimental lattice parameters [22-27]. Some studies [23,24] provide justifications for using experimental lattice parameters. Erba et. al [24] state "*We present here a fully ab initio approach.... As concerns the structure, within the harmonic approximation, here extensively used, no lattice expansion can be predicted, so that the experimental equilibrium lattice parameter at 298 K was used. In principle, however, linear thermal expansion coefficients could be computed within the so-called quasiharmonic approximation by minimizing Helmholtz's free energy (computed from the whole phonon dispersion of the crystal) at a given temperature, as a function of the cell volume or,* *alternatively, from the knowledge of the isothermal bulk modulus, the mode Grüneisen parameters and the mode contributions to the specific heat.*" The underlying principle is that using experimental lattice parameters in first principles calculations is justified since they can be validated by separate *ab initio* studies.

<u>We can extend this very same principle to using experimental DWF.</u> Currently, there are several *ab initio* methods [21, 25-31] to calculate DWF from first principles. In these studies, ab initio DWF has already been calculated for several materials including many semiconductors [21,29]. It is most important to note that *ab initio* DWF closely match the experimental values [21, 25-31]. <u>Hence, we can assume it to be a general feature that ab initio methods give accurate DWFs</u>. This justifies using experimental DWF in ab initio electronic structure calculations to obtain finite temperature band structures, charge densities etc.

For materials where ab initio DWF already exist [21, 25-31] they can be readily used with the above formalism. For materials where ab initio DWF do not exist, it would be preferable to simultaneously perform such a (separate) study. This is because even though experimental DWF can be used, it will eventually have to be validated by ab initio studies.

## 3.2 Quasi *ab initio* formalism

We have used the analogy of experimental lattice parameters to justify using experimental DWF in ab initio studies. However, there are important differences. Finite temperature ab initio studies (even using experimental lattice parameters) usually combine Density Functional Theory with lattice dynamics and also give dynamical information. However, using the finite temperature

DWF corrected potential in electronic structure calculations gives results that are thermal averages. That is, no dynamical information can be obtained in this formalism.

It is clear that the present formalism lacks the full power or possibilities of true *ab initio* approaches. Hence, we call it the Quasi Ab Initio formalism. It is Quasi since it only gives thermal-averaged values and no dynamical information can be obtained. It is Ab Initio if the potential is constructed from first principles atomic calculations.

**3.3 Example**

For psuedopotentials, currently, no examples of ab initio studies exist despite Allen's proof [2] that DWF can be included from the beginning. This was unfortunately due to a wrong impression. As discussed above, the results will be ab initio even when experimental DWF are used with ab initio pseudopotentials. It can be easily implemented in existing computational methods.

In this article, we have extended Allen's result for pseudopotentials to crystal potentials. It is useful to present an example. It is of great interest to know that *such a study already exists in published literature* where it is described as a "practical means" to calculate finite temperature band structures. We refer to the finite temperature band structure of NiS calculated more than forty years ago by Kasowski [6] using the Linear Combination of Muffin Tin Orbitals (LCMTO) method developed by Andersen and Kasowski [32].

In this study [6], the core potential was initially constructed from first principles atomic charge densities, making it an ab initio potential. As a "practical means" an experimental DWF was used to modify this potential. As discussed above, even in this case the finite temperature potential must be considered to be ab initio potential. Next, the core part of the potential was replaced by a smooth function for computational convenience. As seen in sec-2, using a DWF implies that this formalism cannot be used in all-electron methods. Only valence electron energies and wavefunctions can be obtained by this formalism. This justifies the replacement of the core part of the potential by a smooth function in Kasowski's study [6]. Using this potential, the band structure was calculated.

The effect of thermal vibrations on the band structure of NiS is found to be [6] *"The bands which have the DWF incorporated are different in that overlap of the S p bands with the Ni d bands is increased... examination of the wavefunctions indicates that the hybridization with the p states increases for the DWF case... overlap of the d and p bands is at least an order of magnitude more sensitive to thermal vibrations than to discontinuous lattice expansion."*

Kasowski's results [6] are incomplete as they ignore the self-energy terms, since at that time only either one of these two terms were incorporated. It follows from Allen's proof [2] that Kasowski's results can be improved by calculating the self-energy corrections from the finite temperature wavefunctions and energies obtained.

However, as concerns the first DWF step, *the present work transforms (forty years later) Kasowski's empirical results [6] into ab initio results by providing a theoretical formalism. It appears to be the first and only ab initio finite temperature band structure study.*

## 4. Relevance of the Quasi *Ab Initio* approach

The great necessity of the present formalism becomes evident when the current status of finite temperature *ab initio* calculations is considered. We consider two of the most important features of interest in an ab initio calculation, namely 1) band structures and 2) charge densities.

For semiconductors, the temperature dependence of band gaps has been well known for 50 years. For important semiconductors, finite temperature ab initio studies have been performed [9, 13-17, 22] to study band gaps. Yet, these studies do not report finite temperature band structures. To the best of our knowledge, even recent finite temperature band structures are based on empirical approaches [7].

Advances in Convergent Beam Electron Diffraction (CBED) [33-37] have led to accurate measurements of valence electron charge densities. The experimental bonding charge densities or more correctly, their Fourier Transforms, the structure factors, are obtained at finite temperatures in CBED.

Yet, *ab initio* studies to model these results fall broadly into two categories. Firstly, the experimental results are reduced to the (T = 0 K) static lattice case [33,34] by factoring out the DWF in the experimentally measured structure factor. For example, in a recent (2011) study on the bonding electron density in Aluminum, Nakashima et. al [33] state *"All structure factors were converted to T = 0 K and all charge densities presented here are for T = 0 K."*

Alternately, empirical methods [36,37] are used to obtain calculated finite temperature charge densities or structure factors. In these studies [36,37], electronic structure calculations were first performed for the T = 0 K static lattice case. The charge density was partitioned into separate muffin-tin spheres and in the region between them. Next, different DWF were assigned to each of these regions to finally obtain finite temperature structure factors. This method is clearly empirical and no theoretical justification was provided.

It is clear from the above examples that, at present, <u>no simple</u> *theoretically justified* finite temperature ab initio formalism exists to obtain the two most important quantities of an electronic structure calculation, band structures and charge densities (or wavefunctions).

It is clear that the Quasi Ab Initio formalism can fulfill the above requirements. In particular, we note the closeness of Kasowski's approach [6] to the above mentioned [36,37] approach. Till now, both were empirical approaches. The present work has transformed the former into an ab initio approach. The latter method [36,37] becomes an ab initio method if the DWF corrections are applied to the muffin-tin potential rather than to the calculated charge densities.

The above discussion highlights the great potential of the Quasi Ab Initio formalism. If only equilibrium properties are required at finite temperatures, then it is the preferred method. When this formalism is inadequate, especially when dynamical information is needed, then the computationally intensive [22, 24] fully *ab initio* methods are necessary.

## 5. Conclusion

Thermal vibrations cause the atoms to vibrate and alter the external potential. Long ago, Allen (Phys. Rev. B, 18 (1978), p.5217) proved that at finite temperatures the pseudopotential form factors are corrected by a Debye-Waller Factor (DWF). We have generalized this result to the crystal potential consisting of the nuclear and core electron potential. The generalization to the nuclear potential alone fails due to the breakdown of the rigid-atom approximation. This formalism gives equilibrium properties and no dynamical information can be obtained and, is hence, referred to as the Quasi Ab Initio formalism. Analogous to the use of experimental lattice parameters in ab initio studies experimental DWF can also be used. The justification is identical; the experimental parameters can be validated by separate ab initio studies. Our work transforms, forty years later, Kasowski's empirical study (Phys. Rev. B, 8 (1973), p.1378) into the first (and possibly only) ab initio finite temperature band structure calculation. This formalism opens the way to obtain ab initio finite temperature thermal-averaged properties from a single calculation.

# Table 1

Silicon atom energies at various temperatures obtained from Eq.4 by modifying the code RAtom [19]. The mean-squared vibration amplitudes are ab initio calculated values from Table 6 of Ref.21. (* static lattice energy values match NIST reference data [20])

.

| Temp (K) | $\langle u_i^2(T) \rangle$ (Å$^2$) | E Tot (Ha) | E 1s$^2$ (Ha) | E 2s$^2$ (Ha) | E 2p$^6$ (Ha) |
|---|---|---|---|---|---|
| 0* | 0 | -288.1983966 | -65.1844261 | -5.0750558 | -3.5149382 |
| 0.001 | 0.002471 | -221.7084580 | -38.0569774 | -3.3267008 | -3.9720857 |
| 100 | 0.003196 | -214.5276748 | -35.3591088 | -3.1539349 | -4.0167161 |
| 200 | 0.004865 | -202.2466933 | -30.9626472 | -2.8738056 | -4.0762212 |
| 400 | 0.008772 | -184.0485948 | -25.0566186 | -2.4990150 | -4.1033874 |